\documentclass[12pt]{article}

\usepackage[applemac]{inputenc}
\usepackage[T1]{fontenc}
\usepackage{amsmath}
\usepackage{amsfonts}

\usepackage{geometry}                
\usepackage[parfill]{parskip}    
\usepackage{graphicx}
\usepackage{amssymb}
\usepackage{epstopdf}
\usepackage{amsthm} 
\swapnumbers
 
\theoremstyle{definition} 
 
\theoremstyle{remark} 
 
\theoremstyle{plain}

\usepackage[applemac]{inputenc}
\usepackage[T1]{fontenc}                                                                         
\usepackage{amsmath}
\usepackage{amsfonts}

\swapnumbers 
\theoremstyle{plain}

\begin{document}

\title{Majorana equation and its consequences in physics and philosophy}
\author{Daniel Parrochia}
\date{}
\maketitle

\textbf{Abstract}
There will be nothing in this text about Majorana's famous disappearance or about the romantic mythology that surrounds him (see \cite {Kle}). No more on the sociological considerations of the author (see \cite {Maj3}) or on its presumed "transverse" epistemology (see \cite {Syn}). We focus here on the work of the physicist, and more particularly on his 1937 article on the symmetrical theory of the electron and the positron, probably one of the most important theory for contemporary thought. We recall the context of this article (Dirac's relativistic electron wave equation) and analyze how Majorana deduces his own equation from a very general variational principle. After having rewritten Majorana's equation in a more contemporary language, we study its implications in condensed matter physics and their possible applications in quantum computing. Finally, we describe some of the consequences of Majorana's approach to philosophy. \\

\textbf{Physics and Astonomy Classification Scheme (2010)}: 01.65.+g, 01.70.+w.\\

\textbf{key-words} Dirac, Majorana, neutrinos, quasi-particules, supraconduction, quantum computer. \\

\section{Introduction}

One of the most beautiful jewels in theoretical physics is the Dirac relativistic electron wave equation (see \cite {Dir2} and \cite {Dir}), at the origin of the notion of antimatter in the end of the 1920s. This discovery has no doubt been often commented upon and the Dirac equation itself has been the subject of a large literature in the scientific field (see \cite{T ha}). However, as we know, there is another deduction of this equation presented by the Italian physicist Ettore Majorana in 1937,  that has the distinction of leading to purely real solutions where the particles are their own symmetrical. The work of Majorana, long overshadowed, then found again in the 1960s (see \cite {Fra}) and in the 1980s (see \cite{Rec}), has been able to know lately a renewed interest, because of discovery of the weakly massive character of neutrinos. The probable existence of a mass for these wall-pass particles allows them to be considered as "Majorana neutrinos", that is to say, particles that are themselves their own antiparticles. We believe that Majorana's equation deserves a philosophical commentary. On the one hand, it proves that the symmetry imagined by Dirac could be pushed further than he himself had seen, which shows the importance of the great mathematical principles in physics. On the other hand, it poses the problem of the consistency and coordination of heterogeneous data within the same formalism, and in a more precise mode than epistemologists, generally, used to describe with the help of their logical or historical representations of scientific research development. Beyond these methodological views, the Majorana equation is also today at the center of major problems in quantum physics and in current cosmology, to the point that it is probably from the experimental verification of Majorana's hypotheses that we can expect a new light on the universe in which we live and, perhaps, the shift of current science into another form of physics. Finally, the whole process of modern physicists shows the importance of symmetry groups for a rational understanding of nature. In this sense, it should, in our opinion, also inspire the philosopher and the specialist of human sciences.

\section{The Dirac equation}

In 1929, Dirac introduced in quantum mechanics a famous equation, the relativistic electron wave equation, which revealed the power of formalisms and gave birth to a whole new world, since it introduced a mathematical representation of the spin of the electron and of the material-anti-matter symmetry. We already had the opportunity to explain the genesis of this equation which derives from the need to express the Hamiltonian of the wave function by using a form that is linear for some operators, and thus compatible with quantum mechanics, but which remains, at the same time, invariant for the Lorentz transformation. This causes Dirac to express a quadratic form like the square of a linear form (see \cite{Par2}). We resume here most of our presentation in \cite{Par1}.

As we know, the Dirac wave equation for the electron (and, more generally, spin 1/2 particles) results from a relativistic Hamiltonian, which, after re-writing by Klein, Gordon and others, was now expressed as:

\[
\{p_{0} - (m^2c^2 + p_{1}^2 + p_{2}^2 + p_{3}^2)^{\frac{1}{2}}\}\Psi =0,
\]
with:

\[
p_{k} = (h/2\pi)\partial/\partial x_{k}.
\]

where $x_{1}, x_{2}, x_{3}$ are the space coordinates of the electron and $x_{0} = ct$. 

After multiplication by the conjugated expression, it comes:
\[
(p_{0}^2 - m^2c^2 - p_{1}^2 - p_{2}^2 - p_{3}^2)\Psi =0.
\]

Moreover, since the wave equation must be linear in $p_ {0}$ (see \cite{Dir}), Dirac introduces $\alpha_ {i} $ coefficients, which allow him to express this one as the square of the linear form:

\begin{equation}
(p_{0} - \alpha_{1}p_{1} - \alpha_{2} p_{2} - \alpha_{3}p_{3} + mc)\Psi= (i\sum_{1}^4 \alpha_{\mu}p_{\mu} - mc)\Psi.
\end{equation}

This is possible, provided that the matrices $\alpha_{\mu}$ satisfy the conditions:

\begin{equation}
\alpha_{\mu}^2 = 1,  \quad  \alpha_{1} \alpha_{2} +  \alpha_{2} \alpha_{1} = 0 \quad \textnormal{for} \mu \ne \nu,
\end{equation}
and:
\begin{equation}
 \beta^2 = m^2c^2, \quad \alpha_{1}\beta +  \beta\alpha_{1} = 0.
  \end{equation}
 
 If we now put $\beta = \alpha_{m} mc$, these relationships can be summed up in the single formula:
 
 \begin{equation}
 \alpha_{\mu} \alpha_{\nu} +  \alpha_{\nu} \alpha_{\mu} = 2\delta_{\mu\nu} \quad (\mu, \nu = 1, 2, 3 \ \textnormal{ou} \ m).
 \end{equation}

If these conditions are satisfied, the first factor of the expression (1) becomes a perfect square and one thus obtains an equivalence between a quadratic form and the square of a linear form. Conditions (2) and (3) mean that $\alpha_{\mu}$ must be taken from a matrix algebra, which is actually the Clifford's $Cl_{1,3}$ algebra of the Minkowski space.

We know that Dirac easily obtains a representation of these matrices $\alpha_{\mu}$ as products of two sets of Hermitian matrices: matrices $\sigma$ and matrices $\rho$, which also makes matrices $\alpha_{\mu}$ Hermitian matrices. Dirac chooses complex matrices, since, according to the knowledge of the time, for the equation to have a meaning and the electrons to be distinct from their antiparticles (positrons), $\Psi$ must be a complex field. Knowing that the matrices $\sigma_{i}^{(P)}$ of Pauli are defined by:

$
\qquad \qquad \qquad
\sigma_{1}^{(P)} =
\begin{pmatrix} 
0 & 1 \\ 
1 & 0
\end{pmatrix}
$
$
\qquad \quad \sigma_{2}^{(P)} =
\begin{pmatrix} 
0 & -i \\
i & 0
\end{pmatrix}
$
$
\qquad \quad \sigma_{3}^{(P)} =
\begin{pmatrix} 
1 & 0 \\
0 &  -1
\end{pmatrix}
$
\\

and that the identity matrix I$_ {2}$ is: 

$$
\textnormal{I}_{2} =
\begin{pmatrix} 
1 & 0 \\ 
0 & 1
\end{pmatrix}
$$

the Dirac matrices are then constructed as matrices 4$\times$ 4: 

\[
\sigma_{i} = \textnormal{I}_{2} \otimes  \sigma_{1}^{(P)}, 
\]
\[
\rho_{i} =  \sigma_{1}^{(P)} \otimes \textnormal{I}_{2}.
\]

where A $\otimes$ B is the Kronecker product\footnote{The Kronecker product, also called "direct product", allows to multiply matrices of different dimensions. For example, given a $m \times n$ matrix A and a $p \times q$ matrix B, the product C = A $ \otimes$ B is the matrix $mp \times nq$ whose elements are defined by $c _{\alpha \beta} = a_{ij} b_ {kl}$, where $ a = p (i-1) + k $ and $ b = q (j-1) + l $. For example, the direct product A $\otimes$ B of two 2 $\times$ 2 matrices such as:
$
A =
\begin{pmatrix}
a_{11} & a_{12} \\
a_{21} & a_{22}
\end{pmatrix}
$
$
\textnormal{et} \ B =
\begin{pmatrix}
b_{11} & b_{12} \\
b_{21} & b_{22}
\end{pmatrix}
$
is the matrix 4 $\times$ 4:
$
C =
\begin{pmatrix}
a_{11}b_{11} & a_{11}b_{12} & a_{12}b_{11} & a_{12}b_{12} \\
a_{11}b_{21} & a_{11}b_{22} & a_{12}b_{21} & a_{12}b_{22} \\
a_{21}b_{11} & a_{21}b_{12} & a_{22}b_{11} & a_{22}b_{12} \\
a_{21}b_{21} & a_{21}b_{22} & a_{22}b_{21} & a_{22}b_{22} 
\end{pmatrix}
$.

Thus, by making Kronecker product of the identity matrix and Pauli matrices, it is easy to find all of Dirac $\sigma$ and $\rho$ matrices.}.

As can easily be verified, these matrices explicitly take the following forms:\\ 

$
\qquad 
 \sigma_{1} =
\begin{pmatrix} 
0 & 1 & 0 & 0 \\
1 & 0 & 0 & 0 \\
0 & 0 & 0 & 1 \\
0 & 0 & 1 & 0 
\end{pmatrix}
$
$
\qquad \quad \sigma_{2} =
\begin{pmatrix} 
0 & -i & 0 & 0 \\
i & 0 & 0 & 0 \\
0 & 0 & 0 & -i \\
0 & 0 & i & 0 
\end{pmatrix}
$
$
\qquad \quad \sigma_{3} =
\begin{pmatrix} 
1 & 0 & 0 & 0 \\
0 & -1 & 0 & 0 \\
0 & 0 & 1 & 0 \\
0 & 0 & 0 & -1 
\end{pmatrix}
$

$
\qquad
 \rho_{1} =
\begin{pmatrix} 
0 & 0 & 1 & 0 \\
0 & 0 & 0 & 1 \\
1 & 0 & 0 & 0 \\
0 & 1 & 0 & 0 
\end{pmatrix}
$
$
\qquad \quad \rho_{2} =
\begin{pmatrix} 
0 & 0 & -i & 0 \\
0 & 0 & 0 & -i \\
i & 0 & 0 & 0 \\
0 & i & 0 & 0 
\end{pmatrix}
$
$
\qquad \quad \rho_{3} =
\begin{pmatrix} 
1 & 0 & 0 & 0 \\
0 & 1 & 0 & 0 \\
0 & 0 & -1 & 0 \\
0 & 0 & 0 & -1 
\end{pmatrix}
$
\\ 

Among other possibilities, Dirac then chooses combinations of matrices:

\begin{equation}
\alpha_{1} = \rho_{1}\sigma_{1}, \qquad \alpha_{2} = \rho_{1}\sigma_{2}, \qquad \alpha_{3} = \rho_{1}\sigma_{3}, \qquad \alpha_{m} = \rho_{3},
\end{equation}

the $\alpha$'s obviously having to satisfy relations (2) and (3).

\section{The Majorana equation}

In his famous 1937 article (see \cite{Maj1} and \cite{Maj2}), Majorana begins by refusing the hypothesis of the "Dirac sea"\footnote{We know that this hypothesis supposes that the quantum vacuum should be conceived as a "sea" of electrons of infinite depth and in which each of them would occupy a level of pure energy, ranging from negative infinity to a certain maximum value corresponding to the level of the sea. In this configuration, any additional energy input, for example, in the form of a photon, would momentarily create a pair of positron-electron respectively corresponding to a empty state in the negative energy, that is to say a "hole" in the sea, and to a filled state of positive energy. A "hole" in the negative energy of the Dirac Sea, that is, an absence of negative energy, would then correspond to a filled state of positive energy, and vice versa. On the problems raised by the hypothesis of the "Dirac Sea", see \cite{Sau}.}, and more specifically the idea that any particle must have an antiparticle different from itself, which can be problematic, precisely, for "neutral" particles such as neutrino. Majorana immediately remarks on the fundamental flaw of Dirac's approach, which consisted in arriving at a symmetric theory with \textit{ad hoc} procedures from an asymmetrical initial situation, instead of systematically exploring, from the starting point, all possible forms of symmetry. One consequence is that Dirac, more a physicist than a mathematician, has, so to speak, "forgotten" a fundamental symmetry, the symmetry of a particle in relation to itself. Majorana will find this one from an entirely new approach to the Dirac equation. \\
	
	The very elegant Majorana point of view consists in showing that one can deduce the Dirac equation from a more fundamental and symmetrical principle than the Klein-Gordon equation or the Hamiltonian of the wave function, i.e. a "variational principle".

Recall that a variational principle is, at the most general level, a physical principle presenting a natural phenomenon as a constrained optimization problem. Examples abound in physics from the Fermat principle to Feynman's principle, passing through Maupertuis' principle of least action, and his generalization by Euler, Lagrange, and Hamilton (see \cite{Bas}, \cite{Fey} et \cite {You}). 
	 
	In this case, Majorana, in an approach that generalizes that of Jordan and Wigner (see \cite{Jor}), intends to start from a principle that describes the quantification procedure for matter waves (spin particles 1/2), allowing not only to give a completely symmetrical form to the theory of the electron and the positron, but especially to construct an entirely new theory for neutral particles (and, of course, for neutrinos)\footnote{In 1937, Majorana thought that his theory could also concern the neutron. But we now know that the neutron ($n \rightarrow u + d + d $), which is composed of a quark \textit{up} and two quarks \textit{down} has an antineutron, ($\bar{n} \rightarrow \bar{u} + \bar {d} + \bar{d}$) composed of the corresponding antiquarks. The quark \textit{up} has a $+ \frac{2}{3}$ charge and the quark \textit{down} a $ - \frac{1}{3}$ charge. With antiquarks that have equal but opposite charges, the total charge of neutrons and antineutrons remains zero. But this is not true of all neutral particles. For example, bosons such as photon (spin-1) or graviton (spin-2) are created by fields that are their own complex conjugate, that is, which are such that $\phi = \phi^*$.}. \\
	
	Since a physical system $S$ is supposed to be described by real variables $q_{1}, q_{2}, ..., q_{n}$, Hermitian symmetric matrices, Majorana defines the Lagrange function $L$ of the system:	
	\begin{equation}
	L = i \sum_{r, s} (A_{rs} q_{r}\dot{q}_{s} + B_{rs} q_{r}q_{s}),
	\end{equation}
	
and he poses the very general variational principle:

\begin{equation}
\delta \int Ldt = 0.
\end{equation}

It should be understood that the quantities $A_{rs}$ and $B_{rs}$ are ordinary real numbers, the former being constant, the second possibly time-dependent, each satisfying the relations:

\begin{equation}
A_{rs} = A_{sr}, \qquad  B_{rs} = -B_{sr}.
\end{equation}

Majorana then shows that, if the principle (6) is trivially always satisfied when the variables commute, in the case of non-commuting variables, the Hermitian matrix of the system must disappear at any instant for arbitrary variations of $\delta q$, which automatically leads to the cancellation of certain equations of motion which can themselves be derived from the Hamiltonian of the system:

\begin{equation}
H = -\sum_{r, s} B_{rs}q_{r}q_{s}.
\end{equation}

After calculation, and when $A$ is reduced to its diagonal shape, Majorana then reaches a simple condition for the quantities $q_{r}, q_{s}$, which must satisfy:
 
 \begin{equation}
 q_{r}q_{s} + q_{s}q_{r} = \frac{h}{4\pi a_{r}} \delta_{rs}.
 \end{equation}
that is, an equation close to equation (4).

The rest of Majorana's talk is better understood if we go back to Dirac's theory. In his book (see \cite{Dir}, 257), this one shows that, thanks to the matrices (5), we can eliminate the imaginary quantities in the wave equation (1), so that in the absence of an external electromagnetic field, it may be rewritten in vector notation:

\begin{equation}
\{p_{0} - \rho_{1}(\bold{\sigma, p}) - \rho_{3} mc\}\Psi = 0.
\end{equation}

Using a number of substitutions in (11), Majorana rewrites the equation as:

\begin{equation}
\{\frac{W}{c} + (\alpha, p) + \beta mc)\}\Psi = 0.
\end{equation}

He then redefined the Dirac matrices as follows:

\begin{equation}
\alpha_{1} = \rho_{1}\sigma_{1}, \qquad \alpha_{2} = \rho_{3}, \qquad \alpha_{3} = \rho_{1}\sigma_{3}, \qquad \beta = -\rho_{1} \sigma_{2}.
\end{equation}

To make the expression (11) real, he proposes to divide the equations (12) by the quantity $- \frac{h}{2 \pi i}$ and introduce the following changes of variables:
\[
\beta' = -i \beta, \qquad \mu = \frac{2\pi mc}{h}.
\]
He then gets the expression (14):

\begin{equation}
\{\frac{1}{c}\frac{\partial}{\partial t} - (\alpha, \textnormal{grad}) + \beta' \mu \}\Psi = 0.
\end{equation}

	It follows from this reasoning that (11) splits into two subsets of equations, one concerning the real part, the other the imaginary part of the wave function $\Psi$. By putting $\Psi = U + iV$ and then considering the subset of real equations as acting exclusively on $U$, Majorana shows that these equations can be precisely derived from the very general variational principle previously posed by him, which gives them a more natural foundation than in Dirac's theory. Indeed, the equation:
	
	\begin{equation}
\{\frac{1}{c}\frac{\partial}{\partial t} - (\alpha, \textnormal{grad}) + \beta' \mu \}U = 0.
	\end{equation}
	
which is in fact the equation (14) considered as acting on $U$, can be deduced from the variational principle:
\[
\partial \int i \frac{hc}{2 \pi}U^* \{\frac{1}{c}\frac{\partial}{\partial t} - (\alpha, \textnormal{grad}) + \beta' \mu \}U dq dt = 0.
\]
	
	The conditions (3) are satisfied, since the anti-commutation relations (10) hold. We have, explicitly:
	
\begin{equation}
U_{i}(q)U_{k}(q') + U_{k}(q')U_{i}(q) = \frac{1}{2} \delta_{ik} \delta(q-q'), 
\end{equation}

	According to (9), the energy of the system is expressed by the Hamiltonian:
	
\begin{equation}
H = \int U^* \{-c(\alpha, p) - \beta mc^2\} U dq.
\end{equation}

relations (16) and (17) being obviously invariant for the Lorentz transformation. If we then add to these equations their analogs containing the quantity $V$, with their anticommutation relations, we find the framework applied by Jordan-Wigner to the Dirac equation without external electromagnetic field.

But the major consequence of the real character of the $U$ part is that the variables $a_{r}$ describing the particles $\gamma$ and their conjugates $\bar{a}_{r}$, describing their associated antiparticles, are the same. We have:

\begin{equation}
a_{r}(\gamma) = \bar{a}_{r}(\gamma).
\end{equation}

In other words, Majorana has demonstrated that the neutral particles involved in his deduction, and first and foremost neutrinos, must be their own antiparticles.

\section{Majorana today}

Using the tensorial writing and applying the Einstein convention, we are now rewriting equation (1) in the more condensed form:

\begin{equation}
(i\gamma^\mu\partial_{\mu} - m)\Psi = 0.
\end{equation}
which is Dirac's equation in its modern form (see \cite{Pal}).

Abandoning the Majorana derivation from the variational principle, it is then enough to define complex generating matrices to obtain, in real text, real matrices, which is done in the very simple way:

\[
\bar{\gamma}^{0} = \sigma_{2} \quad \bar{\gamma}^{1} = \sigma_{1} \otimes \textnormal{I}_{2} \quad \bar{\gamma}^{2} = i\sigma_{3} \otimes \textnormal{I}_{2} \quad \bar{\gamma}^{3} = i\sigma_{2} \otimes \textnormal{I}_{2}.
\]

This gives, in explicit writing (see \cite{Wil}):\\

$
\qquad \qquad \qquad \qquad
 \bar{\gamma}^{0} =
\begin{pmatrix} 
0 & 0 & 0 & -i \\
0 & 0 & -i & 0 \\
0 & i & 0 & 0 \\
i & 0 & 0 & 0 
\end{pmatrix}
$
$
\quad
 \bar{\gamma}^{1} = 
\begin{pmatrix} 
0 & 0 & i & 0 \\
0 & 0 & 0 & i \\
i & 0 & 0 & 0 \\
0 & i & 0 & 0 
\end{pmatrix}
$

\qquad \qquad \qquad \qquad \qquad \qquad \qquad \qquad \qquad \qquad \qquad \\

$
\qquad \qquad \qquad \qquad
 \bar{\gamma}^{2} = 
\begin{pmatrix} 
i & 0 & 0 & 0 \\
0 & i & 0 & 0 \\
0 & 0 & -i & 0 \\
0 & 0 & 0 & -i 
\end{pmatrix}
$
$
\quad \bar{\gamma}^{3} = 
\begin{pmatrix} 
0 & 0 & 0 & -i \\
0 & 0 & i & 0 \\
0 & i & 0 & 0 \\
-i & 0 & 0 & 0 
\end{pmatrix}
$
\\

	Since these matrices $\gamma$ are multiplied by $i$ in the first member of equation (19), this equation becomes real and only admits real solutions, which corresponds to particles which are their own antiparticles. The derivation of Majorana from a variational principle is forgotten and we retain only the new solutions of the Dirac equation brought by the Italian physicist. We then speak of Dirac's equation "in Majorana's representation", to oppose it to other representations such as, for example, the "chiral" representation of Weyl.

Majorana's recent work has again attracted attention because some of the research we are doing today directly echoes his speculation about these spin 1/2 particles or "Majorana fermions", whose property is to be identical to their antiparticles. Mathematically possible since this article of 1937, their real existence has still not been proven by experience more than seventy years later.

Considerations in the history of physics may suffice to account for the oblivion in which the article of the Italian physicist had fallen. Soon after the discovery of the neutrino in 1956, it appeared that the behavior of neutrinos were very different from that of antineutrinos. For example, according to the conservation law of the leptonic number, the muonic neutrinos $\nu_{\mu}$ emitted during a disintegration of the positive pion $\pi^+$ (since one has: $\pi^+ \rightarrow \mu^+ + \nu_{\mu}$) induce a neutron-proton conversion such that $\nu_{\mu} + n \rightarrow \mu^- + p$, but not a neutron-proton conversion of the type $\nu_{\mu} + p \rightarrow \mu^- + n$. On the other hand, the antineutrinos of the muon $\bar{\ nu}_{\mu}$ emitted during a disintegration of the negative pion $\pi^-$ (with: $\pi^- \rightarrow \mu^- + \bar{\nu}_{\mu}$) satisfy the inverse scheme. It has long been thought that antineutrinos were necessarily other than neutrinos, which explains the lack of interest in Majorana's work.

	Nevertheless, since the discovery of the "flavor" oscillation of neutrinos\footnote{There are, as we know, several types of neutrinos (electronic, muonic and tau). But several experiments have shown a neutrino oscillation during a rather long displacement (for example, the Sun-Earth path), the number of neutrinos of the same species at the arrival being generally deficient compared to the number of the starters, while the neutrinos, which have practically no interaction with matter, have not been able to transform into completely different particles. The oscillation from one species to another is the only possible explanation. In 1998, the Super-Kamiokande experiment made it possible for the first time to highlight this phenomenon. In 2010, researchers at the Opera experiment announced that they had observed an oscillation of the muon neutrino directly to the tau neutrino. Finally, in June 2011, researchers from the T2K project surprised a transformation of the muonic neutrino into an electronic neutrino.}, which shows that the conservation laws of the leptonic number are not, in general, preserved, and that only their sum, at best, can be, the question raised by Majorana was re-examined. The basic fact is that the $\nu$ neutrinos produced in a $\pi^+ \rightarrow \mu^+ + \nu$) decay are in a different state of motion from the antineutrinos $\bar{\ nu}$ issued during a disintegration of the negative pion $\pi^-$ of the type: $\pi^- \rightarrow \mu^- + \bar{\nu}$. The former have a spin oriented to the left while the latter have a right-oriented. Logically, it follows that if neutrinos and antineutrinos are similar, they must have different behaviors in different states of motion.
		
	If we can not observe neutrinos at rest, an index would make it possible to separate the two statements: the double $\beta$ disintegration without neutrino emission, which supposes a violation of the only remaining conservation law, that of the sum of the leptonic numbers\footnote{The double decay $\beta$ is a rare process in which, instead of an electron and a neutrino being emitted, as in classical $\beta$ decay, two electrons and two antineutrinos are actually coming out of the atomic nucleus. This disintegration, which occurs only in certain nuclei such as calcium 48, germanium 76, selenium 82 and some others, occurs only very rarely. In case the neutrinos are Majorana, the two antineutrinos of the $\beta$ disintegration, which are nothing other than neutrinos, should annihilate and disappear from the final state. Several experiments (Supernémo, Gerda, Cuore, Exo, Majorana) track this double disintegration $\beta$ without emission of neutrinos.}. So far, such a situation has never been observed. But it can be, and one is currently seeking, in particular with the germanium and the spontaneous decomposition $\textnormal{Ge}^{76} \rightarrow \textnormal{Se}^{76}$ + 2e, to put in evidence such a situation.

Beyond the neutrino case, the question is whether there are other particles that could be their own antiparticles. Actively researched by particle physicists, at CERN and in many other laboratories, these Majorana particles, if they existed, could be a possible candidate for the solution of the dark matter enigma.

	In recent years, they have also been of interest to condensed matter physicists working on superconducting at high temperatures, who attempt to identify them, not as elementary particles, but as collective excitations in the non-conventional superconducting "gap" (also called "topological gap")\footnote{Recall that the phenomenon of superconducting was discovered in 1911 by a Dutch physicist Heike Kamerlingh Onnes (see \cite{Gav}) and his team, who showed that the electrical resistivity of mercury becomes zero below a certain temperature called critical temperature $T_{c}$, of the order of 4.2 K for mercury. For a long time, it was thought that the superconducting phenomenon could only occur at low temperatures, but in 1986 the discovery of cuprates (associations of atoms of copper, oxygen and other elements) gave hope that one could get superconducting at much higher temperatures. In reality, it is the deeper understanding of the transition to superconductivity (explained in Ginzburg-Landau theory (see \cite{Gin}) as a spontaneous symmetry breaking in a non-electron crystal of conduction, with permanence of conductive edge states linked to a non-trivial topological order) which allowed the most spectacular advance. In 2005, physicists Kane and Mele (see \cite {Kan1}, \cite{Kan2}) suggested that graphene could have such a non-trivial topological order. In graphene, in fact, semi-metal where the conduction and valence bands are touching, taking into account the spin-orbit coupling opens a "gap" between the two bands, so that it becomes insulating. Kane and Moore (see \cite{Kan3}) later generalized this phase to three-dimensional equivalents now called "topological insulators". Associated with superconductors, three-dimensional topological insulators could henceforth produce so-called "topological superconductors", and thus create Majorana fermions. Indeed, just as topological insulators are insulators with robust edge or surface states, topological superconductors have edge or surface states protected by the topology: Majorana fermions.}.
		
	This is how the Quantronic group of the SPEC (Condensed State Physics Department at the CEA) has just launched some experiments to try to detect these "Majorana quasi-particles"\footnote{Quasi-particles, entities similar to particles, have been designed to facilitate the description of particle systems, particularly in condensed matter physics. Among the best known are the "electron holes", which can be thought of as a "lack of electron" in a valence band. But there would be a host of others: Bipolarons (or paired pairs of two polarons), Chargeons (intervening in spin-charge separation situations), Excitons, or bound states of a free electron and a hole, Fluxons (or Quanta of electromagnetic flux), Magnons (or coherent excitations of electron spins in a material), Phonons (or vibratory modes inside a crystalline structure), Plasmons (or coherent excitations of a plasma),     Polarons (or quasi-particles composed of a localized electron coupled with a polarization field), Polaritons (or mixtures of photons and other quasi-particles), Rotons (or states of elementary excitation in helium 4 superfluid), Solitons (or solitary waves propagating without deforming in a non-linear and dispersive medium), and finally Spinons (which intervene in spin-charge separation situations).}.	
	
In the long term, they could still be exploited as information carriers (or "qu-bits") in quantum computers that could become, thanks to this, extraordinarily robust to decoherence. Indeed, particles at the boundary between matter and antimatter, Majorana fermions, in two-dimensional structures inside the solids, would behave like anyons (i.e. neither fermions, or bosons), which topological laws would then make much more resistant to decoherence. The group of the Dutch physicist Leo Kouwenhoven (see \cite{Kou}), of the University of Delft, who tested a hypothesis formulated by the American R. Lutchyn and his collaborators, which predicted that fermions of Majorana should be formed in a magnetic field at the interface of a superconductor and a semiconductor nanotube (see \cite{Lut}), seems to have highlighted the signature of such particles (see \cite{Rei}).

If Majorana fermions have indeed been produced in the nanowire of Dutch researchers and if it is possible to manipulate them, the realization of a quantum computer could be closer than we think. In any case, the extraordinary fertility of the work of the Italian physicist is measured here.

\section{Philosophical considerations}

	We have yet to draw some conclusions from the above.

As we have already had the opportunity to point out on several occasions (see \cite{Par1}, \cite{Par2}), the Dirac equation already revealed, on its own, the power of mathematics. The resolution of the contradiction between the quadratic expression of the relativistic Hamiltonian and the necessary linearity of the wave function of the electron, which leads to expressing a quadratic form like the square of a linear form, made it possible not only to account for this additional degree of freedom of the particles (the spin, formally inscribed in the non-commutative algebra of the matrices associated with the coefficients of the form), but also made arise a whole new world: that of antimatter. Recall that the positron, planned by Dirac, will be discovered by Anderson in 1932.

The formal manipulations of Dirac, however, can not absolutely satisfy a true mathematician, while the existence of negative energy states - the famous "Dirac Sea" - is repugnant to the real physicist. Majorana, who is both at the same time, constructs a theory of "neutral" particles, particularly applicable to the neutrino, predicted by Fermi in 1931, from a particularly elegant deduction. As Etienne Klein rightly notes, "his ideas are so revolutionary that no one could really understand them in the context of the 1930s, especially as they are presented with a perfectly original mathematical formalism, which is based on symmetries abstained that physicists are not yet used to"(see \cite{Kle}, 113).
 How did he reason?
 	
If Bachelard, a French philosopher, author of \textit{The Inductive Value of Relativity} had studied Majorana's approach, he would probably have explained things as follows: from the equation of Dirac, Majorana, by a process of abstraction generalizing, is traced back to the Hamiltonian fundamental schema of this equation, which actually comprises two types of variables $A$ and $B$, those concerning the momentum $p$, and those relating to the mass $m$. These variables are summed and, since it is convenient to describe waves with complex numbers, multiplied by $i$. We thus go back from the Hamiltonian to a function of Lagrange. Two very general conditions will then be posed:

\begin{enumerate}
\item Majorana claims that the derivative of the integral vanishes (see equation (7)), which imposes the existence of an extremum, in this case a minimum corresponding to a lower action of the nature, which supposes the perfectly rational conviction that it is conducted heuristically.
\item Majorana then makes common sense assumptions about the variables $A$ and $B$. Those that are independent of time must remain identical for a permutation of their indices, while the others will obviously see their inverted values for the same permutation (see equation (6)). It is a way of saying that the laws of nature must be independent of the observer.
\end{enumerate}
 
 From these very simple demands are then derived conditions that are very close to those usually associated with the Dirac equation (10). It remains for Majorana to reformulate it in such a way that it leads to real and no more complex solutions, which amounts to finding complex matrices which, multiplied by $i$, will render the Dirac equation real.
	
	But our interpretation is still too empirical. There are, in fact, fundamental mathematical reasons which explain the possibility of Majorana's approach and which, as a result, also show the difference between his point of view and that of Dirac.

This is because the structure of the wave equation is in fact closely related to the Lorentz symmetry of space-time. The fact that the physical laws are supposed - by Einstein as by most physicists - independent of the observer imposes certain symmetries which make it possible to connect the points of view of the different observers. Now these symmetries are precisely expressed in the group of Lorentz (see \cite{Ast}) of the theory of relativity or, if we include the gravitational questions, in the group of Poincaré, also called, moreover, "inhomogeneous Lorentz group".

The Dirac or Majorana wave functions, also called "spinors" and which are linear functions of quantum mechanics, are, on the contrary, related to symmetries expressed in the theory of complex Hilbert spaces. So that the ratio expressed by Dirac between the relativistic Hamiltonian and the wave function is in fact a special case of a much more general mathematical question which is that of the unitary projective representations of the Poincaré group on complex Hilbert spaces, representations studied by many authors - beginning with Wigner (see \cite{Wig}) in 1939 - that is to say, at a time slightly subsequent to Majorana's article. Wigner's work was later taken up and developed on a purely mathematical basis by George Mackey (see \cite{Mac1}, \cite{Mac2} and \cite {Mac3}).
	
	Although the representations of Poincaré's group on \textit{complex} Hilbert spaces were thus studied for a long time, on the other hand, there was, until recently, no systematic study of the representations of this same group on \textit{real} Hilbert spaces, which corresponds to Majorana's point of view. In this context, it appears that the Dirac spinor is an element of a complex vector space of dimension 4, while the Majorana spinor is an element of a real vector space of the same dimension. Moreover, it has been shown (see \cite{Ped}) that the projective representation of the Poincaré group on the Dirac spinel field is anti-unitary and reducible, whereas it is orthogonal and irreducible on the Majorana spinors field for finite masses.

But we can draw even more general lessons from the mathematical physicist's approach. Beyond the simple problem of Majorana, the research and the exploitation of mathematical symmetries deserve to be elevated to the rank of method. This is what George Mackey did (see \cite{Mac4}) in a remarkable article that summarizes this approach as follows:

	Let $S$ be a "space" or a "set" and $G$ a group of automorphisms of $S$, that is, a group of one-to-one transformations of $S$ within itself. Let $s(x)$ be the transform of $s \in S$ by $x \in G$. Ordinarily, $S$ will be provided with a precise structure that will be preserved during these transformations, so that $s \longrightarrow s(x)$ transformations are $S$ symmetries. It will be appropriate to allow members of $G$ other than the identity $e$ to define the identity application so that a group quotient $G / N$ is the real group of transformations.

Now let $F$ be a vector space of complex valued functions in $S$, which is $G$-invariant, in the sense that $s\longrightarrow f(s(x)$, the translated of $f$ by $x$, is in $F$ each time $f$ is in $F$. Then, for each $x$ of $G$, the application $f \longrightarrow g$, where $g(s) = f(s (x)$, is a linear transformation $V_ {x}$ of $F$ in $F$ and $V_{xy} = V_{x} V_{y}$ for all $x$ and all $y$ of $G$. The application $x \longrightarrow V_{x}$ is thus an example of what is called a (linear) \textit{representation} of the group $G$.

More generally, a linear representation of the group $G$ will be, by definition, a $x \longrightarrow homomorphism\ W_{x}$ of $G$ in the group of all bijective linear transformations of a vector space $H(W)$. A common method used in physics is to try to find $M_{\lambda}$ subspaces of the $F$ space such as:
	
\begin{enumerate}
\item $V_{x} (M_{\lambda}) = M_{\lambda}$ for all $x$ and all $\lambda$;
\item Any $f$ element of $F$ is only a finite or infinite sum $f = \sum f_{\lambda}$ where each $f _{\lambda} \in M_{\lambda}$;
\item The subfields $M_{\lambda}$ or else are not likely to be decomposed anymore, or are of simpler structure than $F$. Naturally, we can have a topology on $F$, so that we can give meaning to the idea of "infinite sum". More generally, we can also consider continuous direct sums, or direct integrals, or even functions with complex or vector values. Of course, each $M_{\lambda}$ can be re-represented $V^{\lambda}$ which is called an "under-representation", and we can then speak of a "direct integral decomposition" or a "direct sum decomposition" of $V$. This decomposition, usually, greatly simplifies the starting problem.
\end{enumerate}

A real philosopher would go a little further. Philosophy being supposed to be knowledge of all aspects of reality, a philosopher worthy of the name should ask whether it might be possible in this discipline to start from a space or a sufficiently large set to encompass all or most aspects of this reality, and to define, from there, a group of $G$ automorphisms large enough to express the symmetries one can discover. Then, he should ask if there are (linear) representations of this group and if these, in some cases, can be unitary or irreducible representations of the group $G$. It is probable that such a method then makes it possible to better pose certain philosophical problems, and, in any case, to relate them, as well as their possible solutions, to the general conditions which made their appearance possible and which must justify a certain plausibility. We would thus avoid defining many impossible philosophical universes and so many perfectly inconsistent theories. More than seventy years after Majorana, it does not seem that philosophers have taken the measure of the methods that physics has put in place and which have allowed it to make such remarkable progress. But if, in the field of philosophy and the human sciences, we managed to put in place the program that has just been sketched in broad strokes, no doubt, then, that we could finally give ourselves the means to advance a little these disciplines that remain, for now, a poor relative of our rationality.


\begin{thebibliography}{}

\bibitem{Ast}
Aste, A., «A direct road to Majorana fields», {\it Symmetry} 2 (2010) 1776–1809.

\bibitem{Bas}
Basdevant, J.-L., {\it Le principe de moindre action et les principes variationnels en physique}, Paris, Editions Vuibert, 2010.

\bibitem{Dir2}
Dirac, P. A. M., «The Quantum Theory of the Electron», {\it Proc. R. Soc. Lond.} A 1,  vol. 117, No 778, (1928), 610-624

\bibitem{Dir}
Dirac, P. A. M., {\it The Principles of Quantum Mechanics} (1930), Oxford, Clarendon Press, 1989.

\bibitem{Fey}
Feynman, R. P., Leighton, R. B., et Sands, M., {\it The Feynman Lectures on Physics}, New York, Addison-Wesley,1964.

\bibitem{Fra}
Fradkin, D. M., «Comments on a Paper by Majorana Concerning Elementary Particles», {\it  American Journal of physics}, Volume 34, Issue 4 (1966), 314-318. Republié dans {\it Electronic Journal of Theoretical Physics (EJTP)}, 3, No. 10 (2006), 305–314.

\bibitem{Gav}
Gavroglou K and Goudaroulis Y. (ed.), {\it Kamerlingh Onnes, Heike, Through measurement to knowledge : the selected papers of Heike Kamerlingh Onnes (1853-1926)}. Dordrecht, Boston, Kluwer Academic Publishers, 1991.

\bibitem{Gin}
Ginzburg V. L. and Landau, L. D., {\it Zh. Eksp. Teor. Fiz.}  20, 1064 (1950), trad. angl. dans: L. D. Landau, {\it Collected papers}, Oxford, Pergamon Press, 1965, p. 546.

\bibitem{Jor}
Jordan, P., Wigner, E., «Über das Paulische Äquivalenzverbot», {\it Zeitschrift für Physik}, 47, No 9 (1928), 631-651.

\bibitem{Kan1}
Kane, C. L., «Topological Insulator: An Insulator with a Twist». {\it Nature} 4 (5)(2008), 348.

\bibitem{Kan2}
Kane, C. L., Mele, E. J. (30 September 2005). «Z2 Topological Order and the Quantum Spin Hall Effect», {\it Physical Review Letters} 95 (14), 2005.

\bibitem{Kan3}
Kane, C. L., Moore, J. E., «Topological Insulators», {\it Physics World} 24 (2011), 32.

\bibitem{Kle}
Klein, E., {\it En cherchant Majorana, le physicien absolu}, Paris, Flammarion, 2013.

\bibitem{Kou}
Mourik1, V., Zuo, K., Frolov S. M., Plissard, S. R., Bakkers, E. P. A. M., Kouwenhoven, L. P., 
«Signatures of Majorana Fermions in Hybrid Superconductor-Semiconductor Nanowire Devices», {\it Science}, Vol. 336 No. 6084 (2012), 1003-1007 

\bibitem{Lut}
Lutchyn, R. M., Sau, J. D., and Das Sarma, S., «Majorana Fermions and a Topological Phase Transition in Semiconductor-Superconductor Heterostructures», {\it Phys. Rev. Lett.} 105, 077001.

\bibitem{Mac1}
Mackey,  G. W., «Induced representations of locally compact groups», I, {\it Annals of Mathematics}, 55, (1952), 101-139.

\bibitem{Mac2}
Mackey, G. W., «Induced representations of locally compact groups», II, {\it Annals of Mathematics}, 58, (1953), 193-221.

\bibitem{Mac3}
Mackey, G. W., «Infinite dimensional group representations», {\it Bulletin of the American Mathematical Society}, 69, (1963) 628-686.

\bibitem{Mac4}
Mackey, G. W., «Harmonic Analysis as the Exploitation of Symmetry - A historical survey», {\it Bulletin
 (New Series) of the American Mathematical Society}, Volume 3, No 1 (1980), 543-698.
 
\bibitem{Maj1}
Majorana, E., «Teoria simmetrica dell’ elettrone e del positrone», {\it Nuovo Cim.} 14 (1937), 171–184.

\bibitem{Maj2}
Majorana, E., «A symmetric theory of electrons and positrons», translated from {\it Il Nuovo Cimento}, vol. 14, 1937, pp. 171-184, by Luciano Maiani in {\it Soryushiron Kenkyu}, 63 (1981) 149-462.

\bibitem{Maj3}
Majorana, E., «Il valore delle leggi statistiche nella fisica e nelle scienze sociali», {\it Scientia}, Quarta serie, vol. 36, Febbraio- Marzo (1942), 58-66; trad. anglaise : E. Majorana, «The value of statistical laws in physics and social sciences», {\it Quantitative Finance} 5, (2005), 133–140.

\bibitem{Pal}
Pal, Palash B., «Dirac, Majorana and Weyl Fermions», {\it Am. J. Phys.}, vol. 79 (2011), 485-498.

\bibitem{Par1}
Parrochia, D., {\it Les grandes révolutions scientifiques du XX$^e$ siècle}, Paris, P.U.F., 1997.

\bibitem{Par2}
Parrochia, D., Micali, A., Anglès, P., {\it L'unification des mathématiques, algèbres géométriques, géométrie algébrique et philosophie de Langlands}, Paris, Hermès-Lavoisier, 2013.

\bibitem{Ped}
Pedro, L., «The Majorana spinor representation of the Poincare group»,  site internet : viXra:1305.0201.

\bibitem{Rec}
Recami, E., {\it Il caso Majorana: epistolario, documenti, testimonianze}, Milan, Oscar Mondadori, 1987 et 1991. Rééd., Presso Di Renzo Editore, Rome, 2008. 

\bibitem{Rei}
Reich, E. S., «A solid case for Majorana fermions», {\it Nature}, 483 (2012), 132.

\bibitem{Sau}
Saunders, S., «The Negative-Energy Sea», in S. Saunders and H. Brown (ed.), {\it The Philosophy of Vacuum}, Oxford, Clarendon Press, 1991, 65-109.

\bibitem{Syn}
{\it Revue de Synthèse}, «Ettore Majorana, de la légende à la Science», Volume 134, No 1, Mars 2013.

\bibitem{Tha}
Thaller, B., {\it The Dirac Equation}, Berlin, Heidelberg, New-York, Springer-Verlag, 1992.

\bibitem{You}
Yourgrau, W. Mandelstam, S., {\it Variational Principles in Dynamics and Quantum Theory}, New York, Dover Publications, 1979.

\bibitem{Wig}
Wigner, E., «On Unitary Representations of the Inhomogeneous Lorentz Group», {\it The Annals of Math- ematics} 40, No 1 (1939), 149.

\bibitem{Wil}
Wilzec, F., «Majorana return», {\it Nature Physics} 5, (2009), 614 - 618. 



\end{thebibliography}
\end{document}